\documentclass[12pt,preprint]{aastex}

\usepackage{graphicx}

\def\deg      {{\ifmmode^\circ\else$^\circ$\fi}}

\begin{document}


\title{A POTENTIAL GALAXY THRESHING SYSTEM IN THE COSMOS FIELD
       \altaffilmark{1}
      }

      \author{
	  S. S. Sasaki    \altaffilmark{2,3},
	  Y. Taniguchi  \altaffilmark{3},
	  N. Scoville   \altaffilmark{4,5},
	  B. Mobasher     \altaffilmark{6},
	  H. Aussel        \altaffilmark{7},
	  D. B. Sanders    \altaffilmark{8},
	  A. Koekemoer     \altaffilmark{6},
	  M. Ajiki        \altaffilmark{2},
	  Y. Komiyama     \altaffilmark{9},
	  S. Miyazaki    \altaffilmark{10},
	  N. Kaifu         \altaffilmark{9},
	  H. Karoji      \altaffilmark{10},
	  S. Okamura    \altaffilmark{11},
	  N. Arimoto       \altaffilmark{9},
	  K. Ohta           \altaffilmark{12},
	  Y. Shioya     \altaffilmark{3},
	  T. Murayama    \altaffilmark{2},
	  T. Nagao         \altaffilmark{9,13},
	  J. Koda            \altaffilmark{4},
	  L. Hainline      \altaffilmark{4},
          A. Renzini       \altaffilmark{14},
          M. Giavalisco    \altaffilmark{6},
          O. LeFevre     \altaffilmark{15},
          C. Impey         \altaffilmark{16},
          M. Elvis        \altaffilmark{17},
          S. Lilly         \altaffilmark{18}, 
          M. Rich           \altaffilmark{19},
          E. Schinnerer     \altaffilmark{20}, \&
	  K. Sheth        \altaffilmark{4,21}
}

\altaffiltext{1}{Based on data collected at 
  Subaru Telescope, which is operated by 
  the National Astronomical Observatory of Japan.}
\altaffiltext{2}{Astronomical Institute, Graduate School of Science,
  Tohoku University, Aramaki, Aoba, Sendai 980-8578, Japan}
\altaffiltext{3}{Physics Department, Graduate School of Science \& 
  Engineering, Ehime University, 2-5 Bunkyo-cho, Matsuyama 790-8577, Japan}
\altaffiltext{4}{California Institute of Technology, MC 105-24, 1200 East
  California Boulevard, Pasadena, CA 91125}
\altaffiltext{5}{Visiting Astronomer, University of Hawaii, 2680 Woodlawn
  Drive,  Honolulu, HI, 96822}
\altaffiltext{6}{Space Telescope Science Institute, 3700 San Martin
  Drive, Baltimore, MD 21218}
\altaffiltext{7}{Service d'Astrophysique, CEA/Saclay, 91191 
  Gif-sur-Yvette, France}
\altaffiltext{8}{Institute for Astronomy, 2680 Woodlawn Drive, 
  University of Hawaii, Honolulu, Hawaii, 96822}
\altaffiltext{9}{National Astronomical Observatory of Japan,
  2-21-1 Osawa, Mitaka, Tokyo 181-8588, Japan}
\altaffiltext{10}{Subaru Telescope, National Astronomical Observatory,
  650 N. A'ohoku Place, Hilo, HI 96720, USA}
\altaffiltext{11}{Department of Astronomy, Graduate School of Science,
  The University of Tokyo, 7-3-1 Hongo, Bunkyo-ku, Tokyo 113-0033, Japan}
\altaffiltext{12}{Department of Astronomy, Graduate School of Science,
  Kyoto University, Kitashirakawa, Sakyo-ku, Kyoto 606-8502, Japan}
\altaffiltext{13}{INAF -- Osservatorio Astrofisico di Arcetri, 
  Largo Enrico Fermi 5, 50125 Firenze, Italy}
\altaffiltext{14}{European Southern Observatory,
  Karl-Schwarzschild-Str. 2, D-85748 Garching, Germany}
\altaffiltext{15}{Laboratoire d'Astrophysique de Marseille, 
  BP 8, Traverse du Siphon, 13376 Marseille Cedex 12, France}
\altaffiltext{16}{Steward Observatory, University of Arizona, 
  933 North Cherry Avenue, Tucson, AZ 85721}
\altaffiltext{17}{Harvard-Smithsonian Center for Astrophysics, 
  60 Garden Street, Cambridge, MA 02138}
\altaffiltext{18}{Department of Physics, Swiss Federal Institute 
  of Technology (ETH-Zurich), CH-8093 Zurich, Switzerland}
\altaffiltext{19}{Department of Physics and Astronomy, 
  University of California, Los Angeles, CA 90095}
\altaffiltext{20}{Max Planck Institut f\"ur Astronomie, 
  K\"onigstuhl 17, Heidelberg, D-69117, Germany}
\altaffiltext{21}{Spitzer Science Center, California Institute 
  of Technology, Pasadena, CA 91125}

\begin{abstract}
We report on the discovery of a new potential galaxy threshing system
in the COSMOS 2 square degree field using the prime-focus camera,
Suprime-Cam, on the 8.2 m Subaru Telescope.
This system consists of a giant elliptical galaxy with
$M_V \approx -21.6$ and a tidally disrupted satellite galaxy with
$M_V \approx -17.7$ at a photometric redshift of $z \approx 0.08$.
This redshift is consistent with the spectroscopic redshift of 0.079
for the giant elliptical galaxy obtained from the Sloan Digital
Sky Survey (SDSS) archive.
The luminosity masses of the two galaxies are
$3.7 \times 10^{12} \cal{M}_{\odot}$ and $3.1 \times 10^{9} \cal{M}_{\odot}$,
respectively. 
The distance between the two galaxies is greater than 100 kpc.
The two tidal tails emanating from the satellite galaxy extend over 150 kpc. 
This system would be the second well-defined galaxy
threshing system found so far. 
\end{abstract}


\section{INTRODUCTION}

Hierarchical clustering
scenarios suggest that present day galaxies assembled from
much smaller building blocks during the course of their evolution
(e.g., Peebles 1993). Recent deep surveys have found small dwarf galaxies
(i.e., building blocks) at $z > 2$ (e.g., Pascarelle, Windhorst, \& Keel 1998; 
Ellis et al. 2001; Taniguchi et al. 2003a; Santos et al. 2004; 
Kneib et al. 2004; for a review see Taniguchi et al. 2003b).
At the present epoch, the assembly process continues
with large galaxies capturing their satellite
galaxies. In order to understand the mass assembly of
galaxies, it is therefore important to investigate such 
minor merger processes in detail.

It is known that most galaxies in the local universe have satellite galaxies
(e.g., Zaritsky et al. 1997 and references therein). Satellite galaxies
are expected to sink to the center of the potential of their host galaxies 
due to dynamical friction
(Ostriker \& Tremaine 1975; Tremaine 1981); however, satellites with long
dynamical friction timescales are still found orbiting their hosts. 
The satellites suffer from tidal disruption as they orbit 
in the galaxy's host dark matter halo. Therefore, the kinematic and
morphological properties of the satellites can be used to investigate the 
structure and potential of the host dark matter halo.

The recent discovery of ultra compact dwarf (UCDs) galaxies
in the Fornax cluster suggests that such tidal disruption processes
can yield small satellite galaxies (Drinkwater et al. 2000a, 2000b; Bekki,
Couch, \& Drinkwater 2001; Bekki et al. 2003). As dwarf galaxies 
orbit around the host galaxy, 
their outer stellar components are tidally removed. 
This process is also referred to as galaxy threshing.
Recently, Forbes et al. (2003) identified the first case of a galaxy
threshing system in an early release image of the Advanced Camera for
Surveys (ACS) on the Hubble Space Telescope (Tran et al. 2003;
de Grijp et al. 2003). The ACS image clearly shows faint, long tidal tails
emanating from a satellite galaxy around an edge-on spiral galaxy
at $z = 0.145$.

Although a pair of tidal tails is often found in major mergers
(Arp 1966; Toomre \& Toomre 1973), it is rare to find such tails in
on-going minor mergers. 
In this respect, the discovery of Forbes et al. (2003) provides 
important information of tidal disruption of satellite galaxies
around the host galaxy. Following Bekki et al. (2001)
we use the term ''galaxy threshing'' for such on-going, 
tidally-disrupted satellite galaxies.

Here, we present the serendipitous discovery of a new potential galaxy
threshing system found in the COSMOS 2-square degree field 
(Scoville et al. 2006).
The most important point is that we find clear evidence for the
disruption of a small satellite galaxy. Destroyed systems are often observed,
but faint tidal remnants are generally too faint to be detected in most cases.
We present the photometric properties of the system and discuss some 
implications.
We adopt a flat universe with $\Omega$$_{\rm matter} = 0.3$,
$\Omega$$_{\Lambda} = 0.7$, and $H_0 = 70$ km s$^{-1}$
Mpc$^{-1}$ throughout this paper, and use the AB system for optical magnitudes.

The Cosmic Evolution Survey (COSMOS) is a treasury program on
the Hubble Space Telescope (HST), awarded a total of 640 HST orbits,
to be carried out in two cycles (320 orbits in cycles 12 and 13 each;
Scoville et al. 2006; Koekemoer et al. 2006).
This is the largest amount of HST time ever, allocated to a single project.
COSMOS is a 2 square degree imaging survey of an equatorial field in
$I$(F814W) band, using the Advanced Camera for Surveys (ACS). 

\section{OBSERVATIONS}

The COSMOS HST survey alone cannot be used to address all scientific
questions without additional ground-based and space-based 
follow-up observations at multiple-wavelength. 
We have been carrying out optical 
multi-band imaging surveys of the COSMOS 2-square degree field 
centered at $\alpha$(J2000) = $10^{\rm h} ~ 00^{\rm m} ~ 28.6^{\rm s}$ and
$\delta$(J2000) = $+02^\circ ~ 12' ~ 21.0''$ (Taniguchi et al. 2006; 
see also Capak et al. 2006) using Suprime-Cam,
which consists of $5\times 2$ CCDs with 2k $\times$ 4k pixels and has 
a pixel scale of $0.''202$ pixel$^{-1}$ (Miyazaki et al. 2002)
on the 8.2m Subaru Telescope (Kaifu et al. 2000).
During our two observing runs in January and February 2004, we obtained
$B$, $V$, $r'$, $i'$, and $z'$ band images of the whole COSMOS field.

The individual CCD data were reduced and
combined using our own data reduction software
(Yagi et al. 2002) and IRAF.
The combined images for individual bands were aligned and
smoothed with Gaussian kernels to match the atmospheric seeing conditions.
The FWHM of the PSF of the final images has been matched to $0".92$
corresponding to the FWHM of the PSF in the $z^\prime$ band image
which had the worst seeing of all the data.
Photometric calibrations are made using spectrophotometric standard stars.
Details of the data reduction processes for the COSMOS ACS data are
given in Koekemoer et al. (2006).

\section{RESULTS}

In Figure 1, we show our five broad-band images of the galaxy threshing
system. The threshing system is a pair of interacting
galaxies labeled A and B in the $B$ band image (Fig. 1).
Galaxy A appears to be a giant elliptical galaxy while galaxy B
appears to be a small satellite disk galaxy. 
It is remarkable that very long tidal tails
are emanating from galaxy B. These features are quite similar
to those found in the first galaxy threshing system reported by 
Forbes et al. (2003).
Although we have not yet obtained optical spectroscopy of galaxies A and B,
their physical association appears to be real, and thus this system is
the second case of galaxy threshing. 
In Figure 2, we show an ACS $F814W$ image of this system. 
In the high resolution HST image, there are many other satellite galaxies
around galaxy A. This group has been identified as ID 10203 in SDSS by 
Merch\'{a}n \& Zandivarez (2005), They are all large, bright
galaxies ($r' \sim 16-17$) shown in Figure 3.
Note that galaxy B is not contained in the SDSS 
catalog, because the spectroscopic redshift of galaxy B was not obtained.

\subsection{Properties of the Two Galaxies in the Threshing System}

In order to study structural properties of the two galaxies in the
threshing system, we examine their surface brightness distributions using
the IRAF "ellipse" program.  
In Figure 4, we show the surface brightness profiles of
galaxy A as a function of $r^{1/4}$. 
It follows a de Vaucouleur's $r^{1/4}$ law,
suggesting that it is an elliptical galaxy, consistent with the SDSS spectrum.
In Figure 5, we show the surface brightness profile of
galaxy B as a function of radius.
It shows an exponential radial profile suggesting that
this galaxy is probably a disk galaxy.

Next, we derive the apparent magnitudes of the two galaxies using 
GALFIT (Peng et al. 2002). 
We apply a de Vaucouleur's $r^{1/4}$ model for galaxy A
and an exponential disk model for galaxy B. 
Since the central region ($r < 1"$) of galaxy A is saturated
in our $i'$ band image, we have masked out this region.
The magnitudes and their errors of both galaxies are summarized in Table 1.
The scale length of galaxy B and the effective radius of galaxy A are
shown in Table 2 and Table 3, respectively. 

We also give results for the Sersic fitting in Table 4. 
The Sersic index of galaxy A ranges from $\approx$ 2.5 to 
$\approx$ 3, being smaller than 4 
(which corresponds to a de Vaucouleur profile).
Note that the central region of galaxy A is saturated in $i'$,
and thus the Sersic index may have large uncertainty.
On the other hand, galaxy B has Sersic index around 1.4
in each filter.
These results seem to be basically consistent with our interpretation that
galaxy A is an elliptical galaxy while galaxy B is a disk galaxy.

At a redshift of 0.08 (corresponding to a luminosity distance of 
363.5 Mpc), we estimate an absolute $V$ magnitude (luminosity) of
$M_V \sim -21.6 ~ (L_V \sim 3.7 \times 10^{10}L_{V,\odot})$ for galaxy A,
and of $M_V \sim -17.7 ~ (L_V \sim 1.0 \times 10^{9}L_{V,\odot})$ for galaxy B.

\subsection{Redshift of the Threshing System}

We found a spectroscopic redshift of $z=0.079$ for galaxy A
(SDSS J100003.2+020146.4) in the SDSS spectroscopic data archive
of third data release (Abazajian et al. 2005).
Unfortunately, no spectroscopic information on the redshift is 
available for galaxy B. 
We derive a photometric redshift for galaxy B using
the $\chi^2$ minimizing method (e.g., Lanzetta, Yahil, \&
Fern\'{a}ndez-Soto 1996).
We generate SED (spectral energy distribution) templates for 
$0 \le z \le 1.0$ with a redshift bin of $\Delta z = 0.01$ using
the population synthesis model GALAXEV (Bruzual \& Charlot 2003).
The SEDs of local galaxies are well reproduced by models whose star-formation
rate declines exponentially ($\tau$ models); i.e., $SFR \propto \exp(-t/\tau)$,
where $t$ is the age of the galaxy and $\tau$ is the time scale of 
star formation.
We adopt $\tau = 1 {\rm Gyr}$ models with a Salpeter initial mass function
(a power index of $x=1.35$ and a stellar mass range of 
$0.01 \le m/M_\odot \le 125$) and solar metallicity, $Z=0.02$.
We then calculate SEDs for ages of $t=1$, 2, 3, 4, and 8 Gyr, corresponding to
the SED of a starburst, Irr, Scd, Sbc, and elliptical galaxy, respectively.
We adopt a visual extinction of $A_V=0$. Note that our threshing 
system consists of basically nearby galaxies. 
Therefore, our results should not be affected by metallicity and $A_V$.
Applying this method, we obtain
a photometric redshift of the galaxy B as $z_{\rm ph}=0.08$,
being consistent with the spectroscopic redshift of the galaxy A.
For the rest of the paper, we adopt $z =0.08$ for the galaxy threshing system
($1 \arcsec \simeq 1.76 \ \rm{kpc}$ at $z = 0.08$).

\subsection{Properties of the Tidal Tails}

We investigate the morphological properties of the two tidal tails
emanating from galaxy B. As shown in Figure 1, 
the two long tidal tails are obvious 
in images of all bands, suggesting that they consist of stellar material.
The lengths of the two tidal tails are 58 arcsec for the
tail extending towards the galaxy A and 30 arcsec for the counter tail 
(east of galaxy B).
These values correspond to lengths of 102 kpc and 52 kpc, 
respectively in the plane of the sky.

The tail extending toward galaxy A appears to be narrower than
the counter tail whose width increases with increasing
distance from galaxy B.
This property is also seen in the first galaxy threshing system found
by Forbes et al. (2003).
In order to study this tendency more quantitatively, 
we examine the surface brightness distributions of the tails.
In Figure 6, we show the $i^\prime$-band surface brightness profile
perpendicular to the tails for the four spatial positions, a, b, c,
and d that are indicated in the right panel of Figure 6.
As shown in the left panel of Figure 6, the tail extending to 
galaxy A is significantly narrower than the counter one.
The tail widths measured at 3$\sigma$ and  5$\sigma$ noise level
for each position are given in Table 5.
The width at position d exceeds 10 kpc even then it is measured at the
5$\sigma$ noise level. However, the width of the tail extending to
galaxy A is only $\simeq$ 6 - 7 kpc. 

There are some knotty structures in the tidal tails. It is likely 
that some of these may be tidal dwarf candidates (e.g., Zwicky 1956; Schweizer 
1978; Barnes \& Hernquist 1992; Yoshida, Taniguchi, \& Murayama 1994; 
Duc \& Mirabel 1998 and references therein).
Future spectroscopic identification will be necessary to confirm them.

We estimate the mean surface brightness  of the tidal tails
in a sky area above $1.5\sigma$ of the surface brightness in each band.
The errors are estimated using following relation;
\[
\sigma^2({\rm{mean\ counts/pixel}}) = \frac{({\rm{mean\ counts/pixel}}) + \sigma^2_{{\rm{background}}}}{N}.
\]
$N$ is the total number of pixels of tidal tail, 
and $\sigma^2_{\rm{background}}$ is $1 \sigma$ sky background noise.
The results are given in Table 6.  
The mean surface brightness is fainter by $\approx$ 1 magnitude than
that of the tail found by Forbes et al. (2003).

We also estimate the total magnitude of the tidal tails in each band.
Errors are estimated using following relation;
\[
\sigma^2({\rm{total\ counts}}) = {\rm{Total\ counts}} + N \times \sigma^2_{{\rm{background}}}.
\]
The results are given in Table 6.
It is found that the magnitude of the tails is comparable to
that of galaxy B in each band. 

In Figure 7, we show the $B - r^\prime$ color distribution in the 
galaxy threshing system. The mean $B - r^\prime$ color of galaxy A
is $1.37 \pm 0.06$, while, that of galaxy B is $1.02 \pm 0.07$. 
There is a small color difference between the 
main body of galaxy B and the two tidal tails.
This suggests that stars located in the tidal tails
were dispersed from the main body of galaxy B,
providing strong evidence that galaxy threshing has been occurring 
in this system.
The total magnitude of the tidal tails is comparable to that of
the main body of galaxy B. We estimated that the luminosity of galaxy B
is $L_V \sim (1.03 \pm 9.48\times10^{-5}) \times 10^9 L_{V,\odot}$,
and that of the tail is 
$L_V \sim (0.94 \pm 2.16\times10^{-3}) \times 10^9 L_{V,\odot}$.
These luminosities are nearly the same. This suggests that
galaxy B lost about half of its 
former mass during the (on-going) threshing event.

\subsection{Dynamical Properties of the Galaxy Threshing System}

We assume that galaxy A has a
mass-to-light ratio of 10 typical for an elliptical galaxies 
and 3 for galaxy B.
Given the $V$-band luminosity in section 3.1, we obtain
$M_{\rm A} \simeq 3.7 \times 10^{12} \cal{M_{\odot}}$ and 
$M_{\rm B} \simeq 3.1 \times 10^{9} \cal{M_{\odot}}$ 
for A and B, respectively.

Next, we estimate the probable pericenter distance ($r_{\rm p}$) for the
galaxy threshing event. 
As shown in Figure 5, galaxy B shows an exponential surface
brightness profile for $r \lesssim 2.9$ arcsec.
If we assume that the mass outside of this radius has been
expelled during the previous encounter between the two galaxies,
and we use the equation of tidal radius as a rough guide of the pericenter
distance, we can then estimate the pericenter as follows,
\[
r_{\rm p} \simeq r \times \left(\frac{2 \cal{M}}{m}\right)^{\frac{1}{3}}.
\]
Using the derived masses of the two galaxies,
we obtain a pericenter distance of $r_{\rm p} \simeq 68$  kpc. 
Bekki et al. (2003) made general numerical simulations of
a galaxy threshing systems with $r_{\rm p} = 65$ kpc.
Mayer et al. (2001) used a similar value, $r_{\rm p} = 75$ kpc.
Our estimate is consistent with the results of 
these numerical simulations.

The projected separation between two galaxies is 102 kpc.
If we adopt an average relative velocity between the two galaxies
of 200 km s$^{-1}$, we find that the previous encounter occurred 
$\tau \sim 5 \times 10^{8}\left(v\ /\ 200 \ \rm{km \ s}^{-1}\right)^{-1}$
yrs ago.

Galaxy B found in this study will merge into galaxy A
after a few billion years after (see Bekki et al. 2001). Even in a
facial phase of such minor mergers, faint remnants could be found around 
the host galaxy. 

Recently, morphological evidence for such a final phase
of minor mergers has been found around  field red elliptical galaxies
(van Dokkum 2005). Deep optical imaging survey for ordinary-looking
galaxies should provide more information on the global, and dynamical 
evolution of galaxies.

\section{Concluding Remarks}

In this paper, we have presented the discovery of a new potential galaxy 
threshing system. Although merging systems and interacting systems are
often observed, it is rare to find faint tidal tails in
on-going minor mergers because their tidal tails are generally 
too faint to be detected. 
Our discovery is important because we find clear 
evidence for the disruption of a small satellite galaxy. 
This is only the second well-defined galaxy threshing system found 
so far.

The HST COSMOS Treasury program was supported through NASA grant
HST-GO-09822. We wish to thank Tony Roman, Denise Taylor, and David
Soderblom for their assistance in planning and scheduling of the 
extensive COSMOS observations.
We gratefully acknowledge the contributions of the entire COSMOS collaboration
consisting of more than 70 scientists. 
More information on the COSMOS survey is available \\ at
{\bf \url{http://www.astro.caltech.edu/~cosmos}}. It is a pleasure the 
acknowledge the excellent services provided by the NASA IPAC/IRSA 
staff (Anastasia Laity, Anastasia Alexov, Bruce Berriman and John Good) 
in providing on-line archive and server capabilities for the COSMOS datasets.
The COSMOS Science meeting in May 2005 was supported in part by 
the NSF through grant OISE-0456439. We would like to thank the 
Subaru Telescope staff for their invaluable assistance. 
IRAF (Image Reduction and Analysis Facility) is 
distributed by the National Optical Astronomy Observatory, 
which is operated by the Association of Universities for Research 
in Astronomy, Inc., under cooperative agreement with the National 
Science Foundation.
This work was financially supported in part by the Ministry
of Education, Culture, Sports, Science, and Technology (Nos. 10044052 
and 10304013), and by the Japan Society for the Promotion of Science 
(15340059 and 17253001). SSS and TN are financially supported by the 
Japan Society for the Promotion of Science (JSPS) through JSPS Research 
Fellowship for Young Scientists.

\clearpage

\begin{deluxetable}{cccccccc}
\tablenum{1}
\scriptsize
\tablecaption{Basic properties of the threshing system. Magnitudes and 
their errors are derived using GALFIT. }
\tablewidth{0pt}
\tablehead{
 \colhead{Name} &
 \colhead{RA(J2000)} &
 \colhead{DEC(J2000)} &
 \colhead{$B$} &
 \colhead{$V$} & 
 \colhead{$r'$} &
 \colhead{$i'$} &
 \colhead{$z'$} 
}
\startdata
J095959.6+020206(B) &  9 59 59.61 & +2 02 06.6  & 20.87  & 20.15 & 19.85 & 19.56 & 19.28\\
& & & \small{$\pm 0.01$} & \small{$\pm 0.01$} & \small{$\pm 0.01$} & \small{$\pm 0.01$} & \small{$\pm 0.01$}\\
\\
J100003.2+020146(A) & 10 00 03.23 & +2 01 46.5  & 17.29 & 16.33 & 15.92 & 15.48 & 15.21\\
& & & \small{$\pm 0.00$} & \small{$\pm 0.00$} & \small{$\pm 0.00$} & \small{$\pm 0.00$} & \small{$\pm 0.00$}\\
\enddata
\end{deluxetable}


\begin{deluxetable}{ccccccc}
\tablenum{2}
\footnotesize
\tablecaption{Scale length of the threshed galaxy (galaxy B). 
  These values are derived using GALFIT.}
\tablewidth{0pt}
\tablehead{
 \colhead{Name} &
 \colhead{} & 
 \colhead{$B$} &
 \colhead{$V$} & 
 \colhead{$r'$} &
 \colhead{$i'$} &
 \colhead{$z'$} 
}
\startdata
J095959.6+020206 & $R_{\rm{s}} $(arcsec)& 0.50 & 0.58 & 0.56 & 0.53 &  0.57\\
 & $R_{\rm{s}}$(kpc) & 0.88 & 1.02 & 0.99 & 0.93 & 1.00\\
\enddata
\end{deluxetable}


\begin{deluxetable}{ccccccc}
\tablenum{3}
\scriptsize
\tablecaption{Effective radius of the threshing galaxy (galaxy A). 
  These values were derived using GALFIT.}
\tablewidth{0pt}
\tablehead{
 \colhead{Name} &
 \colhead{} & 
 \colhead{$B$} &
 \colhead{$V$} & 
 \colhead{$r'$} &
 \colhead{$i'$} &
 \colhead{$z'$}  
}
\startdata
J100003.2+020146 & $R_{\rm{e}} $(arcsec) & 3.29 & 3.00 & 3.20 & 2.34 & 4.12\\
 & $R_{\rm{e}} $(kpc) & 5.97 & 5.29 & 5.64 & 4.12 & 5.16\\
\enddata
\end{deluxetable}


\begin{deluxetable}{cccccc}
\tablenum{4}
\footnotesize
\tablecaption{Sersic parameter of both galaxies. 
  These values are derived using GALFIT. Note that the central region 
  ($r < 1$ arcsec) of galaxy A is saturated in $i'$.}
\tablewidth{0pt}
\tablehead{
 \colhead{Name} &
 \colhead{$B$} &
 \colhead{$V$} & 
 \colhead{$r'$} &
 \colhead{$i'$} &
 \colhead{$z'$}  
}
\startdata
galaxy A  & 2.91 & 2.56 & 2.69 & 1.80 & 2.57\\
galaxy B & 1.41 & 1.33 & 1.46 & 1.47 & 1.50\\
\enddata
\end{deluxetable}


\begin{deluxetable}{cccccc}
\tablenum{5}
\footnotesize
\tablecaption{Width of the tidal tails.
  The positions of the cuts across the tidal tails are indicated in Fig.6.}
\tablewidth{0pt}
\tablehead{
 \colhead{} &
 \colhead{} &
 \colhead{a} &
 \colhead{b} & 
 \colhead{c} &
 \colhead{d} 
}
\startdata
3$\sigma $ & (kpc) & $7.6^{+0.1}_{-0.2}$ & $7.9^{+0.3}_{-0.2}$ & $11.8^{+1.1}_{-1.2}$ & $12.3^{+0.1}_{-0.2}$ \\
& (arcsec) & $4.3^{+0.0}_{-0.1}$ & $4.5^{+0.1}_{-0.2}$ & $6.7^{+0.7}_{-0.7}$ & $7.0^{+0.1}_{-0.1}$ \\
\\
5$\sigma$ & (kpc) &$6.4^{+0.1}_{-0.1}$ & $5.5^{+0.2}_{-0.1}$ & $8.5^{+0.3}_{-0.4}$ & $10.3^{+0.3}_{-0.5}$ \\
& (arcsec) & $3.6^{+0.0}_{-0.0}$ & $3.1^{+0.1}_{-0.1}$ & $4.8^{+0.1}_{-0.2}$ & $5.9^{+0.2}_{-0.3}$ \\
\enddata
\end{deluxetable}


\begin{deluxetable}{cccccc}
\tablenum{6}
\footnotesize
\tablecaption{Properties of the tidal tails.}
\tablewidth{0pt}
\tablehead{
 \colhead{} &
 \colhead{$B$} &
 \colhead{$V$} & 
 \colhead{$r'$} &
 \colhead{$i'$} &
 \colhead{$z'$}\\
}
\startdata
\small{Mean surface brightness (mag/arcsec$^2$)}& \small{27.90} \tiny{$\pm$ 0.08} & \small{27.30} \tiny{$\pm$ 0.04} & \small{27.11} \tiny{$\pm$ 0.07} & \small{26.84} \tiny{$\pm$ 0.04} & \small{26.29} \tiny{$\pm$ 0.05} \\
\small{Total magnitude (mag)} & \small{20.91} \tiny{$\pm$ 0.07} & \small{20.18} \tiny{$\pm$ 0.05} & \small{19.82} \tiny{$\pm$ 0.05} & \small{19.66} \tiny{$\pm$ 0.04} & \small{19.23} \tiny{$\pm$ 0.05} \\
\small{Background (mag/arcsec$^2$)} & \small{29.89} & \small{29.39} & \small{29.39} & \small{28.98} & \small{28.05} 
\enddata
\end{deluxetable}

\clearpage



\clearpage    

\vspace{1cm}
\vspace{0.5cm}

\noindent {\bf Fig. 1} Suprime-Cam images of the threshing system. 
Image size of these figures are $2.3' \times 1.5'$ 
(corresponds to 243 kpc $\times$ 158 kpc at $z \simeq 0.08$).
\begin{figure}[h]
\epsscale{0.70}
\plotone{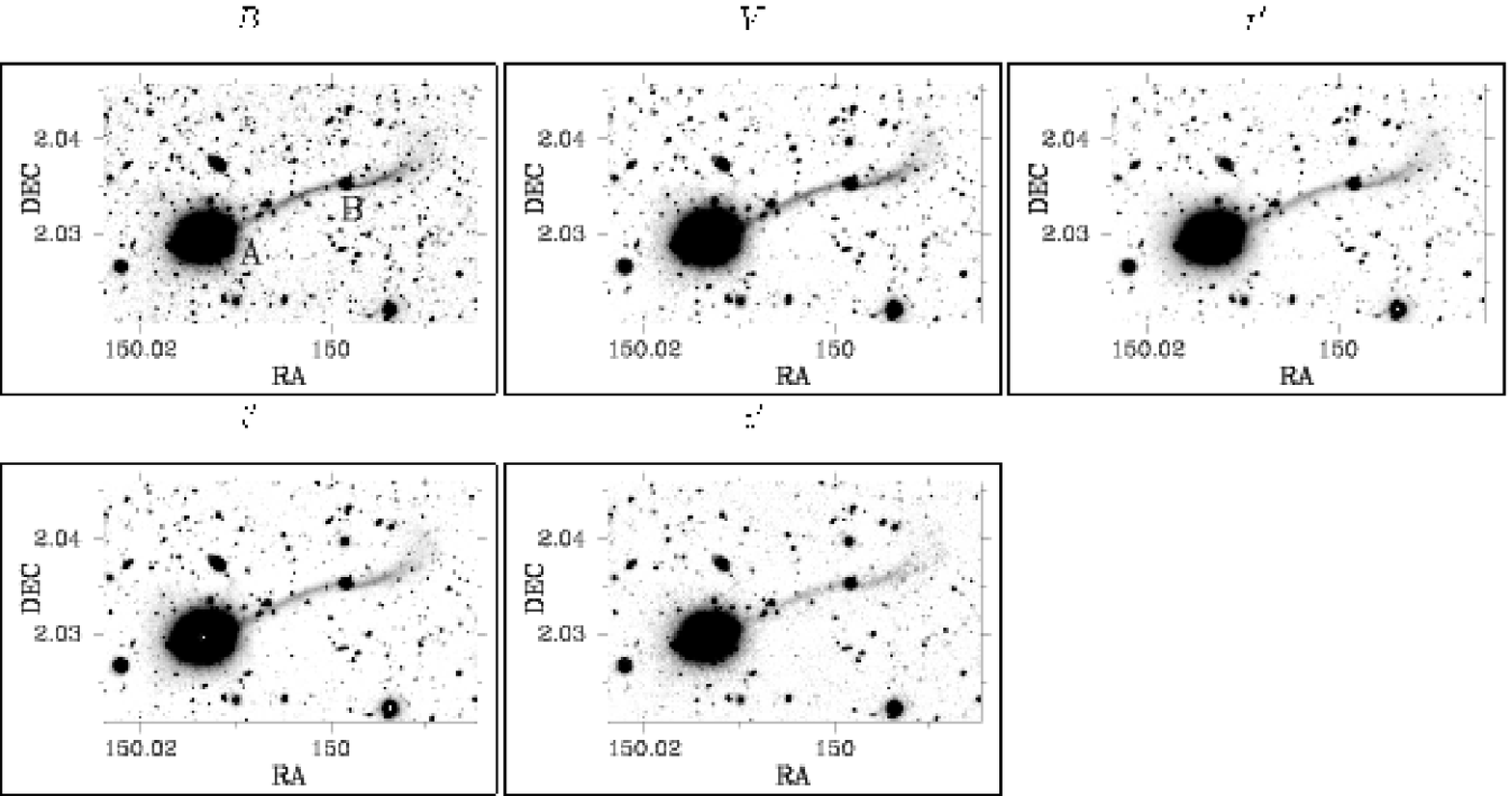}
\end{figure}

\noindent {\bf Fig. 2} HST ACS $I-814$ band image of the galaxy threshing 
system. Image size of this figure is $2.3' \times 1.5'$.
\begin{figure}[ht]
\begin{center}
\includegraphics[width=0.4\textwidth,angle=-90]{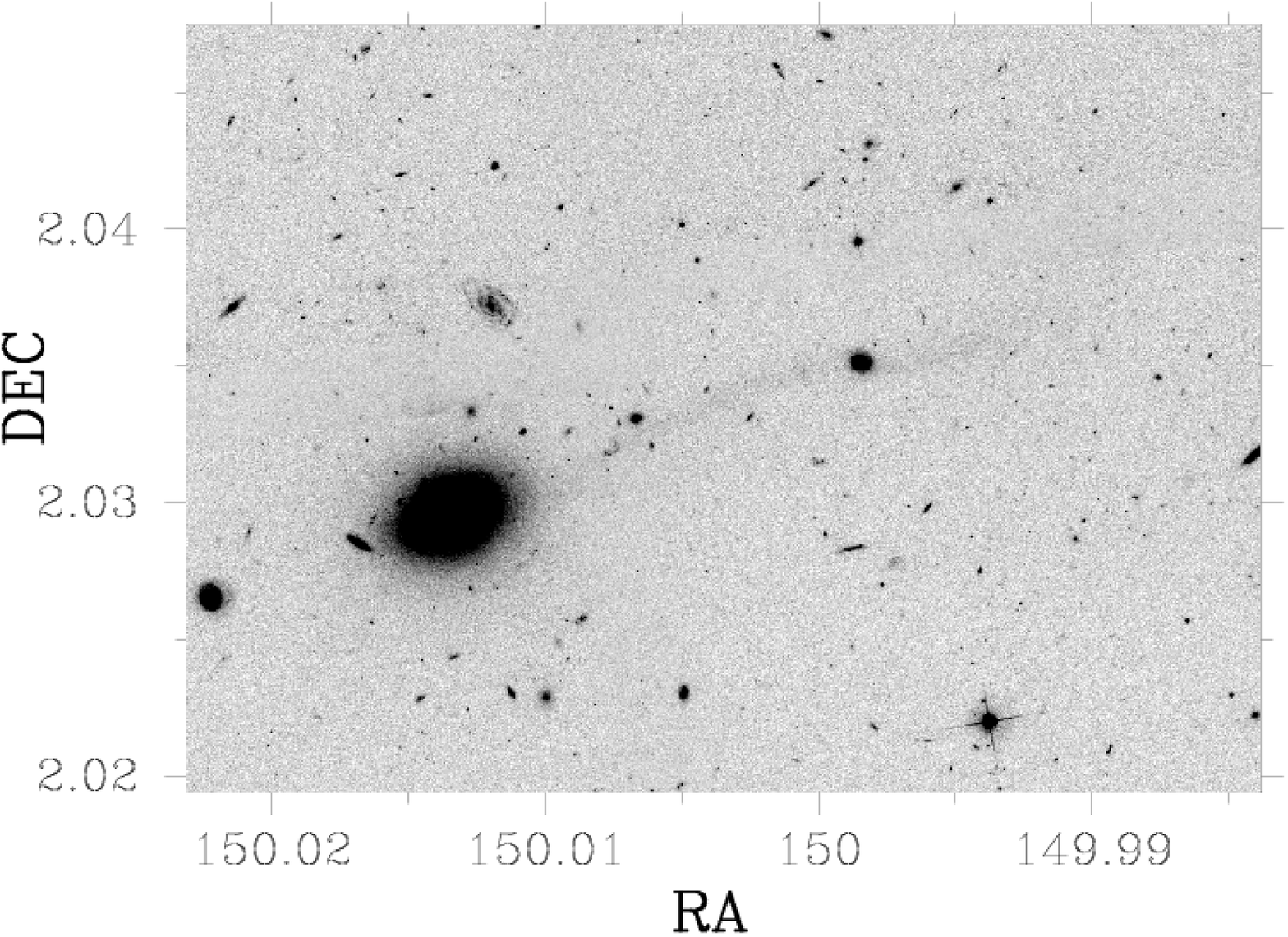}
\end{center}
\end{figure}

\clearpage
\noindent {\bf Fig. 3} Circled galaxies are the member of the galaxy group.
They were identified by Merch\'{a}n \& Zandivarez (2005).
\begin{figure}[ht]
\begin{center}
\includegraphics[width=0.5\textwidth,angle=-90]{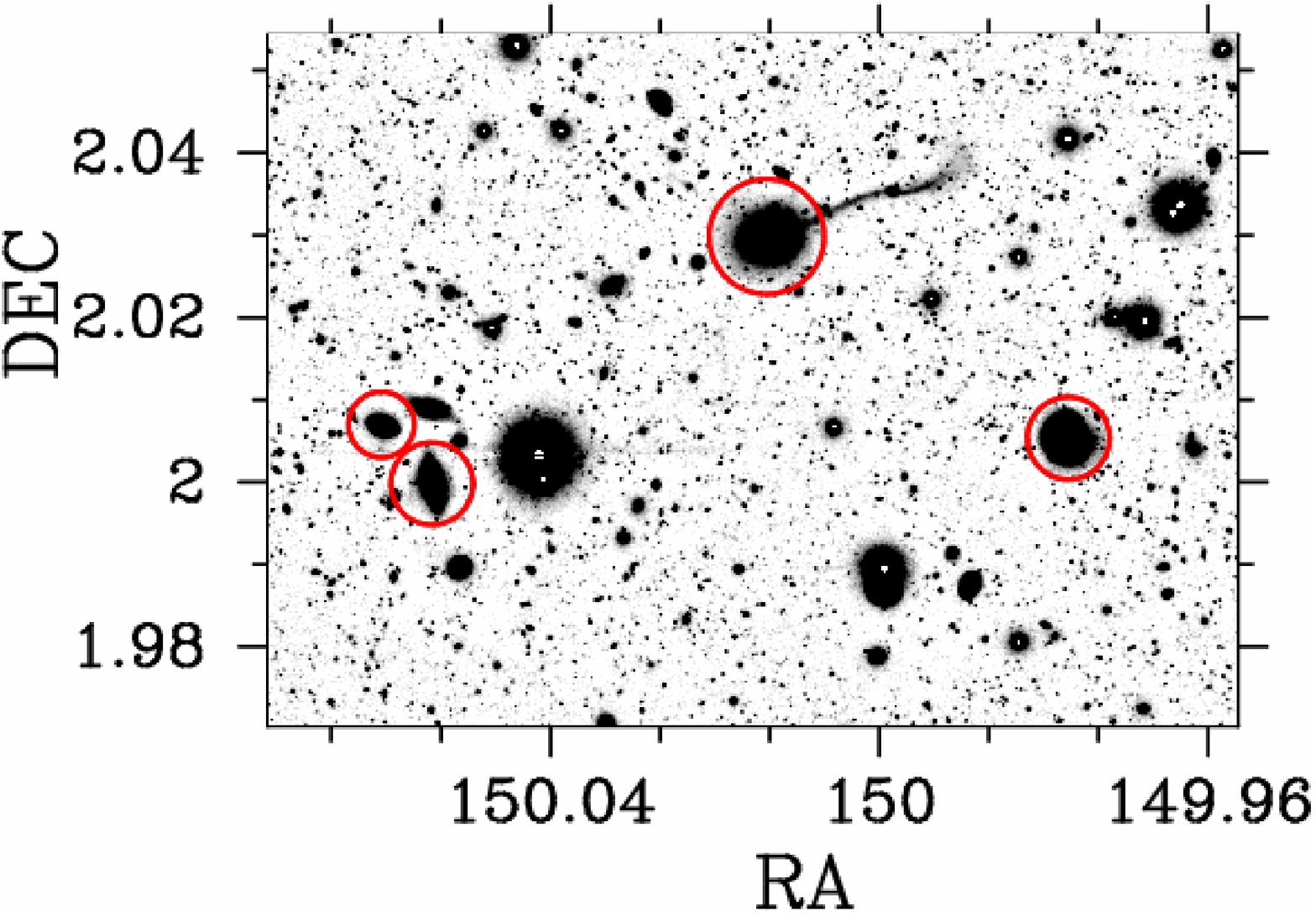}
\end{center}
\end{figure}

\clearpage
\noindent {\bf Fig. 4} Surface brightness profiles of the threshing galaxy. 
Note that the central region ($< 1"$) of the threshing galaxy is saturated 
in $i'$ and has been masked out. We also show the $1 \sigma$ of sky 
background and seeing radius ($= 0.46"$). The values of background are shown 
in the Table 6.
\begin{figure}[ht]
\begin{center}
\includegraphics[width=0.44\textwidth,angle=-90]{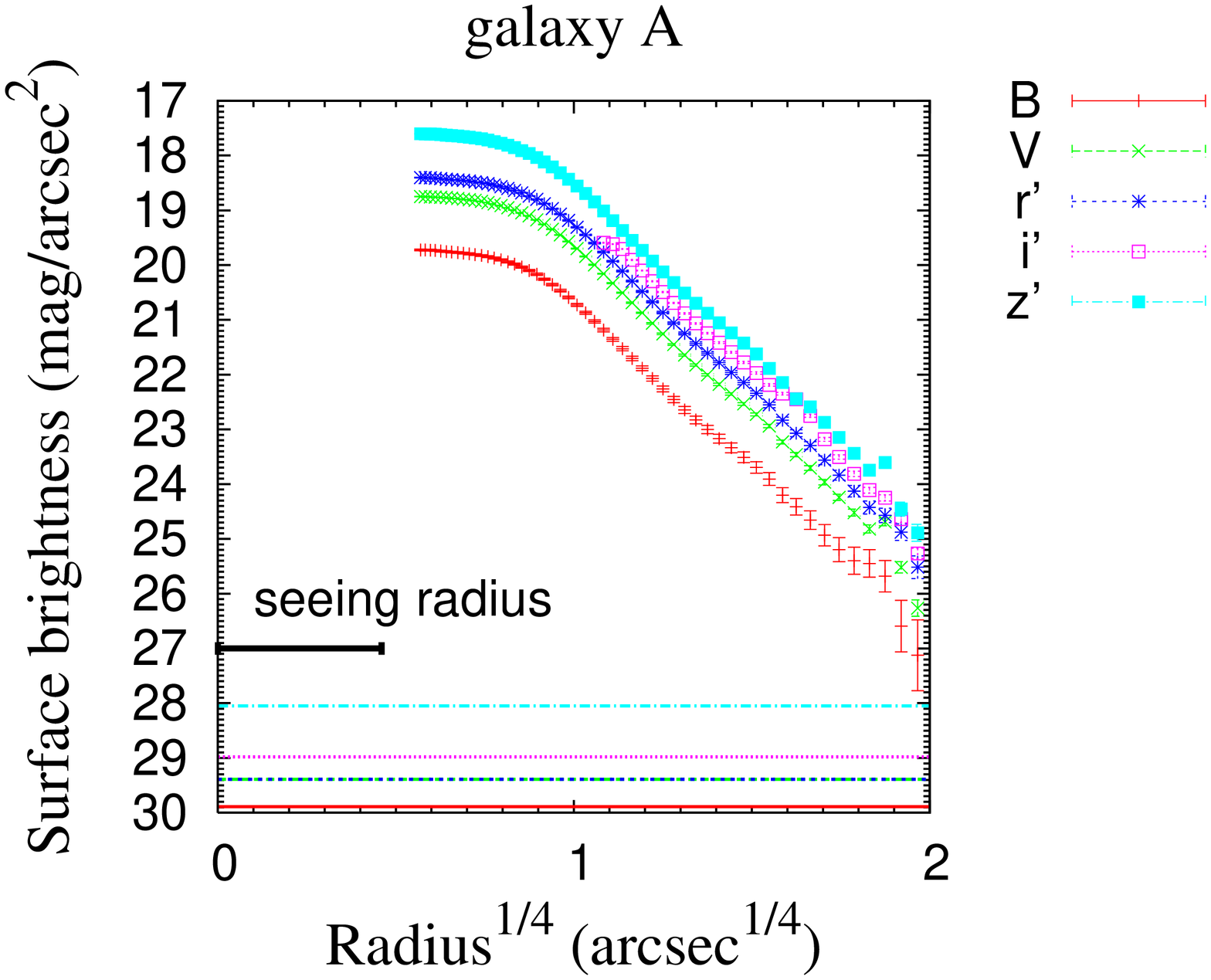}
\end{center}
\end{figure}

\noindent {\bf Fig. 5} Surface brightness profiles of the threshed galaxy. 
We also show the $1 \sigma$ of sky background and seeing radius ($= 0.46"$). 
The values of background are shown in the Table 6.
\begin{figure}[ht]
\begin{center}
\includegraphics[width=0.44\textwidth,angle=-90]{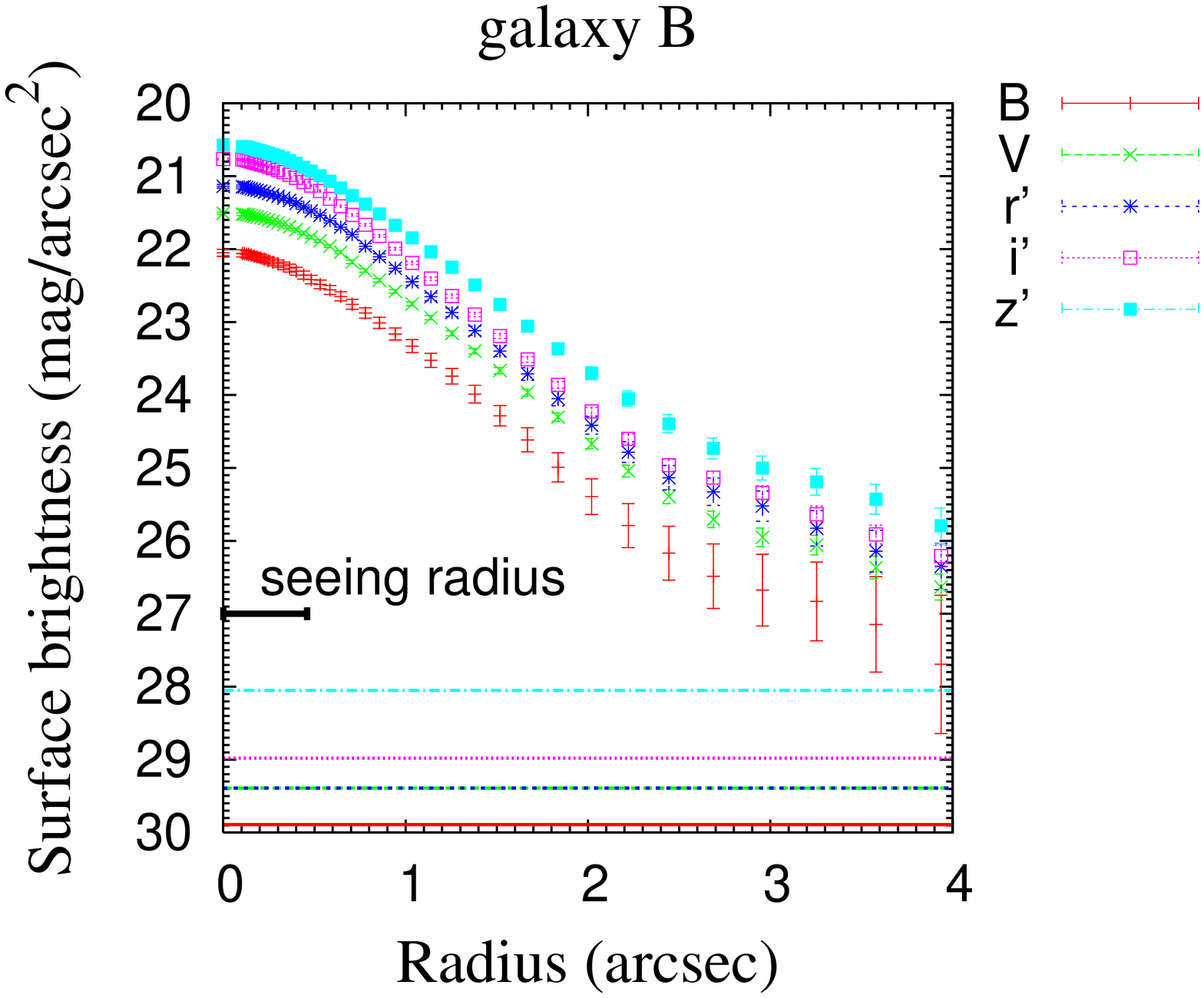}
\end{center}
\end{figure}

\clearpage
\noindent {\bf Fig. 6} Surface brightness distribution perpendicular to the
extent of the tidal tails as a function of position along the tails. 
The investigated spatial positions are marked by bars in the right panel. 
The surface brightness profile at each position is shown in the left panel.
\begin{figure}[htp]
\epsscale{0.40}
\plotone{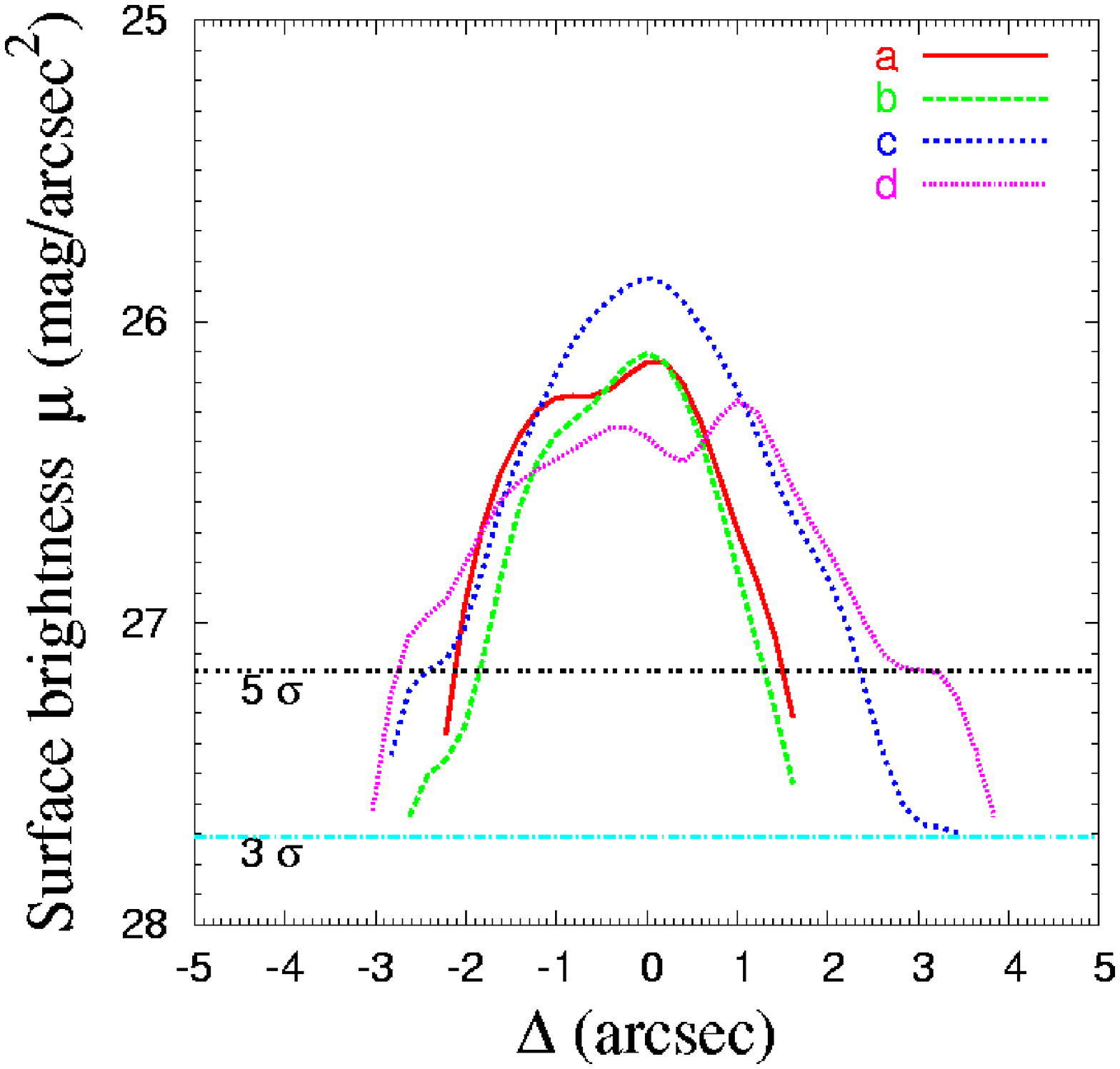}\plotone{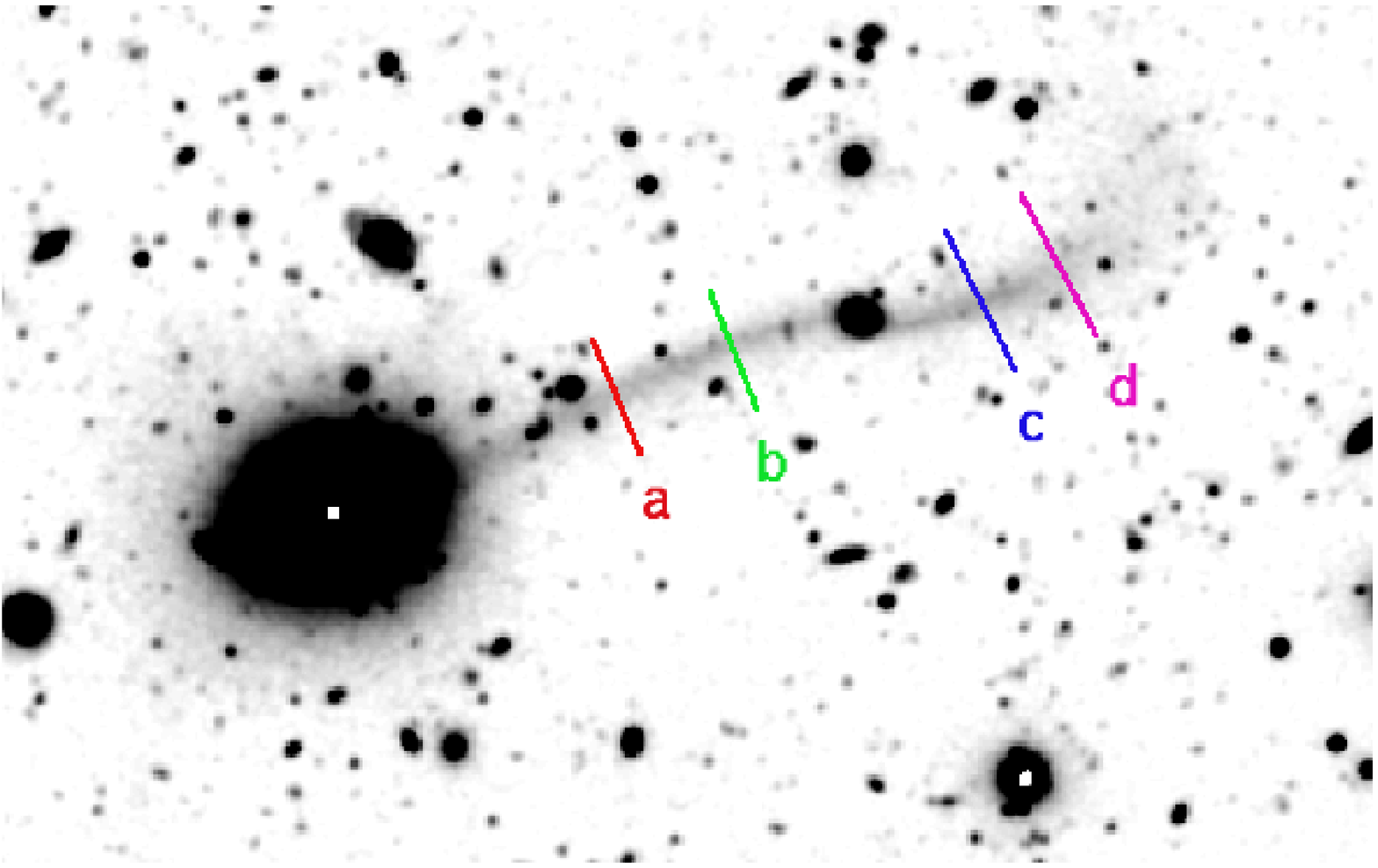}
\end{figure}

\noindent {\bf Fig. 7} $B-r'$ color map.  Image size of this figure is 
$2.3' \times 1.5'$. Outer parts of the threshing system (in particular, 
galaxy A) show redder colors. However, this feature is due to poor 
signal-to-noise ratio because of much lower surface brightness. 
\begin{figure}[ht]
\epsscale{0.5}
\plotone{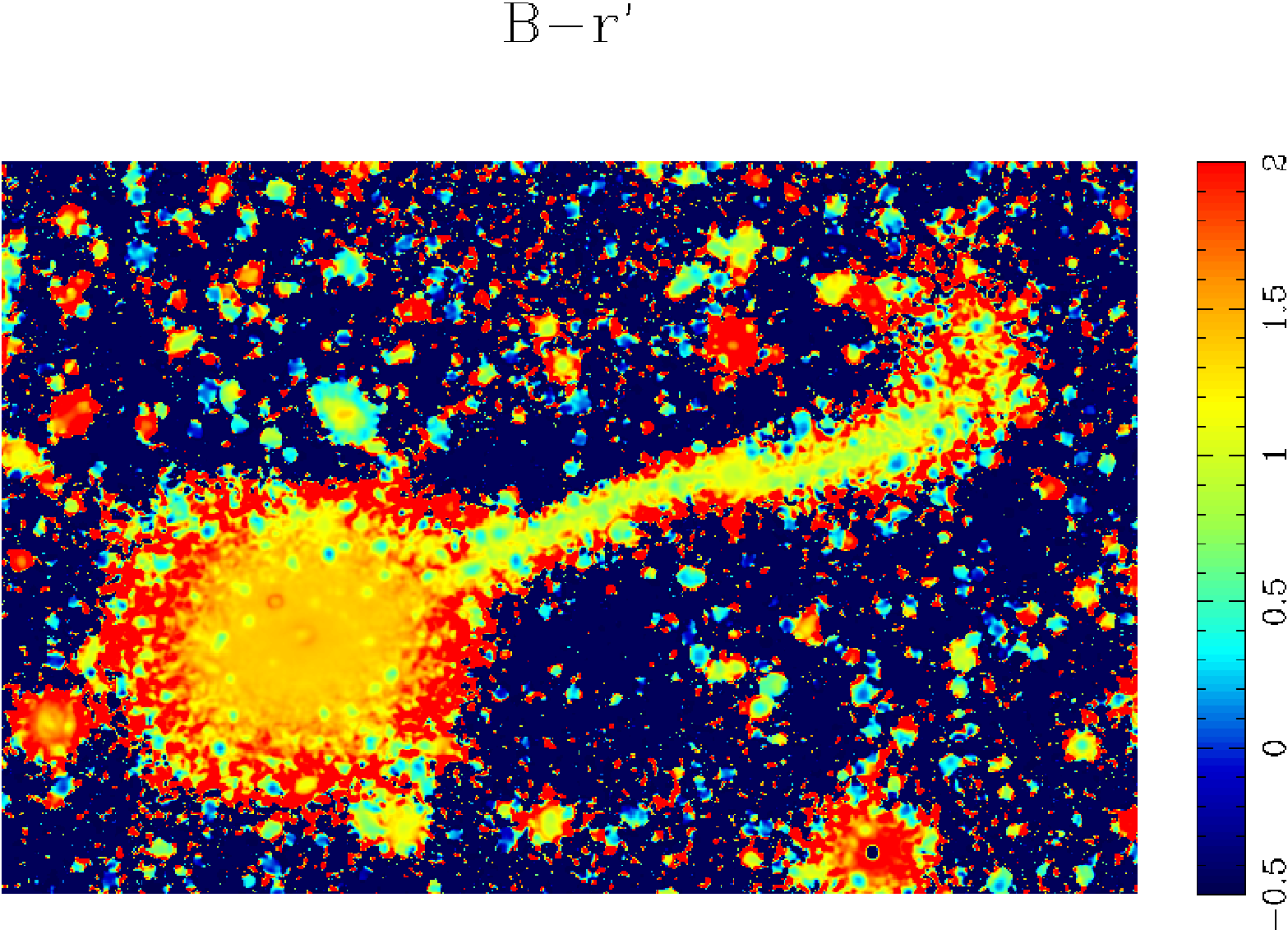}
\end{figure}

\end{document}